\documentclass[journal]{IEEEtran}
\ifCLASSINFOpdf
\else
\fi

\usepackage[T1]{fontenc}
\usepackage[latin1]{inputenc}
\usepackage{graphicx}
\usepackage{amssymb}



\begin{document}

\title{A Controllable Interaction between Two-Level Systems inside a
Josephson Junction }

\author{L. Tian and K. Jacobs}

\maketitle

\begin{abstract}
Two-level system fluctuators (TLS's) in the tunnel barrier of a Josephson junction have recently been demonstrated to cause novel energy splittings in spectroscopic measurements of superconducting phase qubits. With their strong coupling to the Josephson junction and relatively long decoherence times, TLS's can be considered as potential qubits and demonstrate coherent quantum effects. Here, we study the effective interaction between the TLS qubits that is mediated by a Josephson junction resonator driven by an external microwave source. This effective interaction can enable controlled quantum logic gates between the TLS's. Our study can be extended to other superconducting resonators coupling with TLS's.
\end{abstract}


\IEEEpeerreviewmaketitle

\footnotetext[1]{Manuscript was received \today}

\footnotetext[2]{This work was supported in part by the Karel Urbanek Postdoc Fellowship in the Department of Applied Physics at Stanford University.  L. Tian was with the Department of Applied Physics at Stanford University, Stanford, CA 94305 USA. She is at the University of California, Merced, CA 95344 USA (phone: 209-228-4209; e-mail: ltian@ucmerced.edu). }

\footnotetext[3]{K. Jacobs is with the Department of Physics at the University of Massachusetts at Boston, 100 Morrissey Blvd, Boston, MA 02125 USA (phone:617-287-6044; email: kurt.jacobs@umb.edu).}

\section{Introduction}
Two-level system fluctuators (TLS's) are a ubiquitous source of decoherence for solid-state qubits~\cite{DuttaRMP,WeissmanRMP}. In superconducting qubits~\cite{MakhlinRMP}, TLS's have been widely studied both experimentally and theoretically and are often considered as the source of low-frequency ($1/f$) noise~\cite{AstafievPRL, vanHarlingenPRB, WellstoodAPL, MartinPRL, ShnirmanPRL, KochPRL,Paik2008}. In recent experiments, energy splittings in spectroscopic measurements have been observed
in superconducting qubits, showing coherent coupling between the  
TLS's and the qubits~\cite{MartinisPRL, SimmondsPRL}. The TLS's have demonstrated decoherence times much longer than that of the superconducting qubits~\cite{NeeleyNPhys2008} and hence can themselves be considered as effective qubits for testing quantum information protocols~\cite{MartinisPRL, SimmondsPRL,ZagoskinPRL, AshhabNJP, ExpTLS1,ExpTLS2,ExpTLS3,ExpTLS4,ExpTLS5}.

Josephson junctions can be operated as microwave resonators~\cite{BlaisPRA2004,ValenzuelaScience,ChiorescuNature,LupascuNPhys,KochPRL2006,WallraffNature,SillanpaaNature2007,Osborn2007,MajerNature2007,HouckNature2007,MigliorePRB}. In a previous work~\cite{TianPRL2007}, a cavity QED approach~\cite{HoodScience} to characterizing the coupling between TLS's inside a Josephson junction and the junction resonator was suggested where a Jaynes-Cummings model was derived and the coupling between the TLS's and the resonator can be modulated with an applied magnetic field. By measuring microwave transmissions in the junction resonator, various properties of the TLS's can be probed including their spatial distribution and the coupling mechanisms to the junction.

In the following, we study the effective interaction between the TLS's that is mediated by the junction resonator. In our system, because the junction resonator can have a decay rate that is much stronger than that of the TLS's, the effect of the resonator decay needs to be taken into account. We will present the effective interaction both for coupling with a high-Q resonator and for coupling with a strongly damped-resonator.  A microwave source can be applied  to the junction resonator~\cite{TianPRL2007} which provides us with a tool to control the coupling between the TLS's and the resonator. This paper is organized as follows. In Sec. II, we present the model and of the coupling between the TLS's and the junction resonator . In Sec. III, we derive the effective interaction between the TLS's mediated by the resonator mode. In Sec. IV, we will briefly discuss the implementation of quantum logic gates between the TLS qubits and the decoherence effect. We discuss the readout of the TLS qubits in Sec. V and give the conclusions in Sec. VI.

\section{Model}
\begin{figure} 
\begin{center}
\includegraphics[width=8cm,clip]{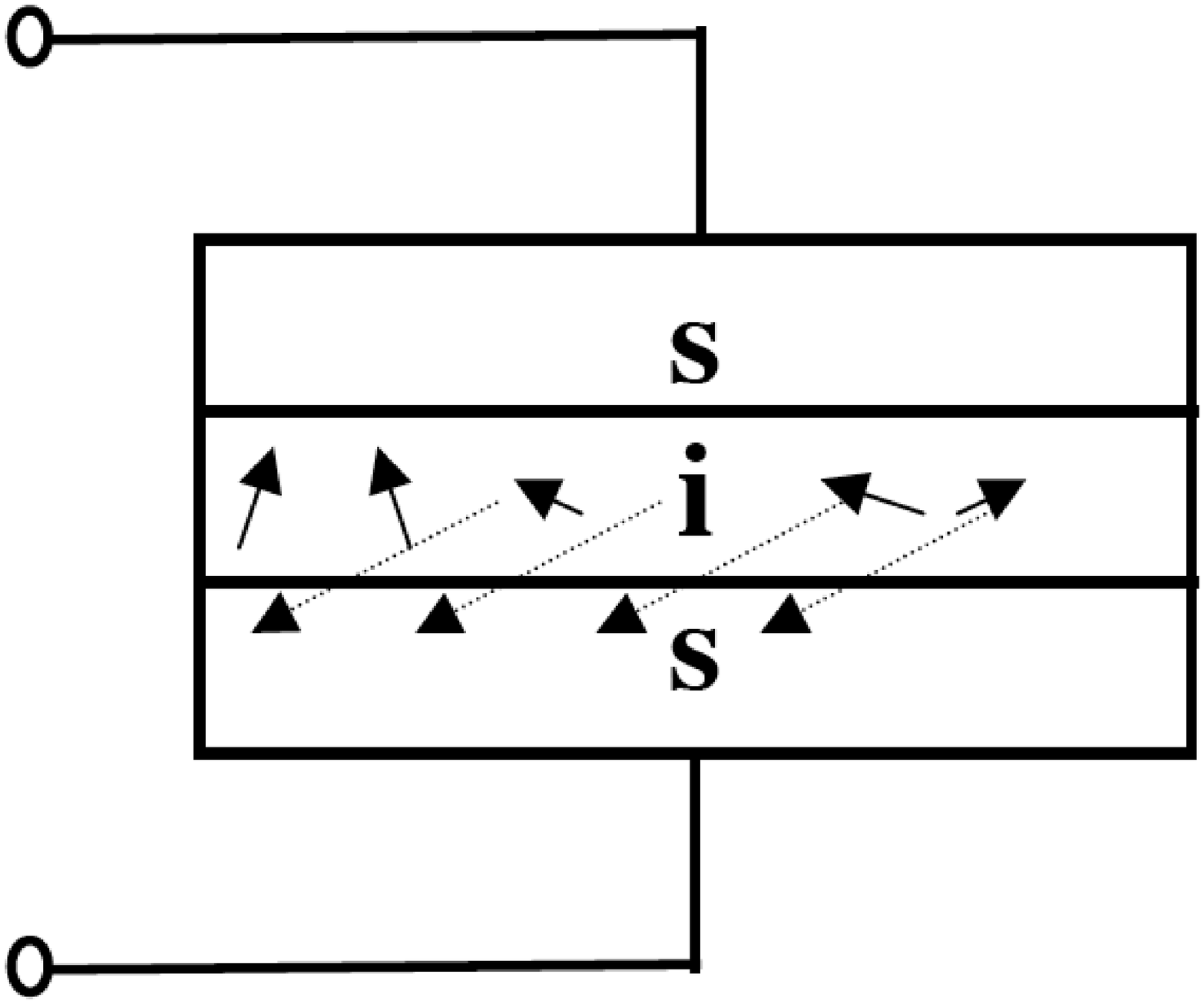}
\end{center}
\caption{Two-level systems (TLS's) inside a Josephson junction. The dashed arrows are
the magnetic field. The letter 'i' indicates the insulating layer and 's' the 
superconducting layers.}
\label{fig1}
\end{figure}

The system is depicted in Fig. \ref{fig1}.  A Josephson junction can be described in terms of the gauge invariance phase $\Phi$ and its conjugate momentum $P_{\Phi}$~\cite{OrlandoBook1991} with a capacitive energy $P_{\Phi}^{2}/2C_{0}$ and a potential energy $-E_{J}\cos(2e\Phi/\hbar)$, where $C_{0}$ is the total capacitance and $E_{J}$ is the Josephson energy. When combined with the  inductance $L$ in an RF SQUID loop, the Hamiltonian of the resonator can be written as 
\begin{equation}
H_{c}=\frac{P_{\Phi}^{2}}{2C_0}-E_{J}\cos(2e\Phi/\hbar)+\frac{(\Phi+\Phi_{ex})^{2}}{2L}
\end{equation}
with external magnetic flux $\Phi_{ex}$ inside the SQUID loop. The Hamiltonian
describes an oscillator mode: $H_{c} \approx P_{\Phi}^{2}/(2C_{0})+C_{0}\omega_{c}^{2}(\Phi-\Phi_{s})^{2}/2$ with a shift $\Phi_{s}$ and a frequency 
\begin{equation}
    \omega_{c}= \sqrt{ \frac{1}{LC_{0}}+\frac{4e^2E_{J}\cos(2e\Phi_s/\hbar)}{\hbar^2C_{0}}}
\end{equation}
both depending on the magnetic flux $\Phi_{ex}$. 

Different coupling mechanisms between TLS's and the junction have been discussed and observed. For example,  TLS's can couple with the critical current of the junction in the form $ -(2e/\hbar)E_{J}\Phi\sum_{n}\vec{j}_{n}\cdot\vec{\sigma}_{n}$ where $\vec{j}_{n}$ and $\vec{\sigma}_{n}$ are, respectively, the polarization vector and the vector of the Pauli spin matrices for the $n^{\mbox{\scriptsize th}}$ TLS. 
Below we assume $\vec{j}_{n}=(j_{x},0,0)$ for simplicity. To the lowest order of $\Phi-\Phi_{s}$ and with the notation $\Phi-\Phi_{s}=\sqrt{\hbar/(2C_{0}\omega_{c})}(a+a^{\dagger})$, the total Hamiltonian of the coupled system is
\begin{eqnarray}
H_{t} & = & \hbar\omega_{c}a^{\dagger}a+H_{i}+\epsilon(t)(a+a^{\dagger})\label{eq:Ht}\\
H_{i} & = & \sum_{n}\frac{\hbar\omega_{n}}{2}\sigma_{nz}+g_{n}(a\sigma_{n+}+a^{\dagger}\sigma_{n-})
\end{eqnarray}
where $a$ ($a^{\dagger}$) is the annihilation (creation) operator of the resonator mode, $\omega_{n}$ is the frequency of the $n^{\mbox{\scriptsize th}}$ TLS, and $g_{n}=E_{J}j_{xn}\sqrt{\hbar/(2C_{0}\omega_{c})}\sin(2e\Phi_s/\hbar)$ is  the coupling constant. The Hamiltonian includes a driving on the resonator mode with a driving amplitude $\epsilon(t)$. Eq. (\ref{eq:Ht}) has the form of a Jaynes-Cummings model that has been widely studied in cavity QED systems. Note that coupling with electrical (dielectric) field inside the junction can be derived similarly~\cite{MartinisPRL,TianPRL2007}.

Dissipative effects can be an important factor when we study the effective coupling between
TLS's.  The decay rate of a junction resonator is much stronger than that of the TLS's. In the following, we treat the resonator decay as a bosonic bath~\cite{WallsBook1994} that couples to the resonator with the Hamiltonian $H_{\kappa}=\sum\hbar\omega_{k}a_{k}^{\dagger}a_{k}+c_{k}(a_{k}^{\dagger}a+a^{\dagger}a_{k})$, where $a_{k}$ ($a_{k}^{\dagger}$) is the annihilation (creation) operator for the bath modes, $\omega_{k}$ is the frequency of the modes, and $c_{k}$ is the coupling constant that is related to the decay rate $\kappa$ with: $\pi\sum _{k}^{2} \delta(\omega-\omega_{k})=\kappa$.  In the master equation approach, the decay can be described in the Lindblad forms as 
\begin{eqnarray}
    \frac{\partial\rho}{\partial t} & = & -i[H_{t},\,\rho]+\kappa\mathcal{L}(a)
                                 \rho. \label{eq:rho}
\end{eqnarray}
In this work, we are interested in TLS's that are in the same frequency range as that of the resonator, i.e. gegahertz, so thermal fluctuations can be neglected. The Lindblad forms can hence be expressed as $\mathcal{L}(o)=\frac{1}{2}(2o\mathcal{L}o^{\dagger}-\mathcal{L}o^{\dagger}o-o^{\dagger}o\mathcal{L})$ for an operator $o$. The intrinsic noise bath of the TLS's is neglected given their long decoherence times.

\section{Effective Interaction}
TLS's do not interact directly due to their low density inside the Josephson junction. However, because of their coupling to the same cavity mode, an effective interaction can be obtained. In this paper, we study the effective interaction in two situations: TLS's coupling with a high-Q resonator with moderate to high decay rate~\cite{TianPreprint}, and TLS's coupling with a strongly-damped resonator where the resonator decay is stronger than the detuning and the coupling between the TLS's and the resonator. 

\emph{High-Q resonator.} We consider the dispersive regime~\cite{BlaisPRA2007,YouPRB,GywatPRB} where the resonator is far detuned from the TLS's with $g_{n}\ll |\Delta_{nc}|$. Here, $\Delta_{nc}\equiv\Delta_{n}-\Delta_{c}$ is the detuning between
the $n^{\mbox{\scriptsize th}}$ TLS and the resonator mode. The effective interaction can be derived by applying a unitary transformation to the Hamiltonian:  $\widetilde{H}_{t}= UH_{t}U^{\dagger}$ where the transformation is
\begin{equation}
U=e^{-\epsilon(a-a^{\dagger})/\Delta_{c}}\prod_{n}e^{-g_{n}(a^{\dagger}\sigma_{n-}-\sigma_{n+}a)/\Delta_{nc}}.
\end{equation}
The total Hamiltonian then becomes $\widetilde{H}_{t}=H_{c}+\widetilde{H}_{eff}+\widetilde{H}_{x}$, where $H_{c}=\hbar\Delta_c a^\dagger a$ denotes the Hamiltonian of the resonator, $\widetilde{H}_{eff}$ the effective Hamiltonian of the TLS's, and $\widetilde{H}_{x}$ the residual coupling between the TLS's and the resonator~\cite{TianPreprint}. We derive 
\begin{equation}
     \widetilde{H}_{eff}=\sum_{n}\left[ \frac{\hbar\bar{\Delta}_{n}}{2}\sigma_{nz}+\frac{\Omega_{nx}}{2}\sigma_{nx} \right] +H_{int}+\widetilde{H}_{k}  
     \label{eq:H1eff}
\end{equation}
which includes single qubit terms with effective detuning $\bar{\Delta}_{n}=\Delta_{n}+(g_{n}^{2}/\Delta_{nc}) (1 - 2\epsilon/\Delta_{c})$ and effective Rabi frequency $\Omega_{nx}=2\epsilon g_{n}/\Delta_{nc}$,  an effective interaction $H_{int}=\sum \lambda_{mn}(\sigma_{n+}\sigma_{m-}+\sigma_{m+}\sigma_{n-})/2$ with the coupling constant 
\begin{equation}
 \lambda_{nm}=\frac{g_{n}g_{m}(\Delta_{nc}+\Delta_{mc})}{2\Delta_{nc}   
                                             \Delta_{mc}},   
   \label{eq:c1}
\end{equation}
and an induced coupling to the bath modes of the resonator $\widetilde{H}_{\kappa}=\sum_{n,k}( g_{n}c_{k}/\Delta_{nc})(\sigma_{n+}a_{k}+a_{k}^{\dagger}\sigma_{n-})$. The residual coupling has the form
\begin{equation}
      \widetilde{H}_{x}=\sum_{n}\frac{g_{n}^{2}}{\Delta_{nc}}\sigma_{nz}
         \left[ a^{\dagger}a+ \epsilon \left(\frac{\Delta_{c} - 2\Delta_{nc}}{2\Delta_{nc} \Delta_{c}}\right)  (a+a^{\dagger}) \right]  \label{eq:Hx}
\end{equation}
with a Stark shift for the resonator and an extra coupling to the resonator amplitude. The unitary  transformation shifts the amplitude of the resonator to zero ($\langle a\rangle \approx 0$) for finite driving $\epsilon$, so that the effect of the Stark shift is always small. 

\emph{Strongly-damped resonator.} The decay rate can reach ten's of megahertz in lossy resonators and cannot be neglected when compared with the coupling between TLS's and the resonator. We set the driving to be $\epsilon(t)=2\epsilon_{0}\cos\omega_{d}t$ with a frequency $\omega_{d}$ and an amplitude $\epsilon_{0}$. To study the dynamics in the Heisenberg picture, we start from Eq. (\ref{eq:Ht}). With $\dot{o}=i[H_{t}+H_{\kappa},\, o]$ for an arbitrary operator $o$, we have
\begin{eqnarray}
\dot{a} & = & -i\omega_{c}a-i\sum g_{n}\sigma_{n-}-i\epsilon-i\sum c_{k}a_{k} \label{dat}\\
\dot{a}_{k} & = & -ic_{k}a-i\omega_{k}a_{k}\label{dak}
\end{eqnarray}
which gives 
\begin{equation}
a_{k}=a_k(0)e^{-i\omega_k t}-i c_{k}\int dt^{\prime}e^{-i\omega_{k}(t-t^{\prime})}a(t^{\prime}),\label{akt}
\end{equation}
with $a_k(0)$ being the noise operator in the Sch\"{o}dinger picture. Substituting Eq. (\ref{akt}) into Eq. (\ref{dat}), we derive the following relation for the resonator mode:
\begin{equation}
\dot{a}=-i\Delta_{c}a-i\sum_{n}g_{n}\sigma_{n-}-i\epsilon_{0}-\kappa a -i \sqrt{\kappa}a_{in}\label{eq:at}
\end{equation}
where $\Delta_{c}=\omega_{c}-\omega_{d}$ is the detuning of the resonator and $a_{in} = (1/\sqrt{\pi})\sum a_k(0) e^{-i\omega_k t}$ is the input field of the bath modes. Similar equations can be derived for the TLS's:
\begin{eqnarray}
\dot{\sigma}_{n-} & = & -i\Delta_{n}\sigma_{n-}+ig_{n}\sigma_{mz}a\nonumber \\
\dot{\sigma}_{n+} & = & +i\Delta_{n}\sigma_{n+}-ig_{n}a^{\dagger}\sigma_{mz}\label{eq:st}\\
\dot{\sigma}_{nz} & = & 2ig_{n}a^{\dagger}\sigma_{n-}-2ig_{n}\sigma_{n+}a\nonumber \end{eqnarray}
where $\Delta_{n}=\omega_{n}-\omega_{d}$ is the detuning of the $n^{\mbox{\scriptsize th}}$ TLS.  In the bad cavity limit with $\kappa\sim \Delta_{n,c}, g_{n}\gg \gamma_{n1},\gamma_{n2}$, we can eliminate the resonator mode by setting the right hand side of Eq. (\ref{eq:at}) to zero. This gives \begin{equation}
  a = -\frac{i\epsilon_{0}+i\sum_{n}g_{n}\sigma_{n-} +i\sqrt{\kappa}a_{in}}{\kappa+i\Delta_{c}} , 
  \label{eq:as}
\end{equation}
where the resonator adiabatically follows the dynamics of the TLS's. The conjugate relation for $a^{\dagger}$ can be derived as well. Now substituting Eq. (\ref{eq:as}) into Eq. (\ref{eq:st}), we can derive a set of equations that govern the dynamics of the TLS's:
\begin{eqnarray}
   \left(\begin{array}{c}
   \dot{\sigma}_{n-}\\
   \dot{\sigma}_{n+}\\
   \dot{\sigma}_{nz}\end{array}\right) & = & A_{n}\left(\begin{array}{c}
   \sigma_{n-}\\
   \sigma_{n+}\\
   \sigma_{nz}\end{array}\right)-\bar{\gamma}_{1}\left(\begin{array}{c}
   0\\
   0\\
   1\end{array}\right)+B_{n}
   \label{eq:sig}
\end{eqnarray}
where $A_{n}$ determines the dynamics of a single TLS, i.e. determines the parameters in the Bloch Equation for a single TLS, with
\[
A_{n}=\left(\begin{array}{ccc}
-i\bar{\Delta}_{n}-\bar{\gamma}_{2} & 0 & i\Omega_{n}+\Lambda_n\\
0 & i\bar{\Delta}_{n}^{\star}-\bar{\gamma}_{2} & -i\Omega_{n}^{\star}+\Lambda_n^\dagger\\
2i\Omega_{n}^{\star}-2\Lambda_n^\dagger & -2i\Omega_{n}-2\Lambda_n & -\bar{\gamma}_{1}\end{array}\right)
\]
and $B_{n}$ determines the effective interaction with 
\[
B_{n}=\sum_{m}\left(\begin{array}{c}
i\lambda_{nm}\sigma_{nz}\sigma_{m-}\\
-i\lambda_{mn}\sigma_{m+}\sigma_{nz}\\
2i\lambda_{mn}\sigma_{m+}\sigma_{n-}-2i\lambda_{nm}\sigma_{n+}\sigma_{m-}\end{array}\right).
\]

The parameters for a single TLS in matrix $A_{n}$ are: the effective detuning $\bar{\Delta}_{n}=\Delta_{n}-\Delta_{c}g_{n}^{2}/(\kappa^{2}+\Delta_{c}^{2})$, the effective Rabi frequency $\Omega_{n}=-ig_{n}\epsilon_{0}/(\kappa+i\Delta_{c})$, the induced dephasing rate $\bar{\gamma}_{2}=g_{n}^{2}\kappa/(\kappa^{2}+\Delta_{c}^{2})$, and an induced decay rate $\bar{\gamma}_{1}=2\bar{\gamma}_{2}$. The induced dephasing (decay) is due to the effective bath
\begin{equation}
  \Lambda_{n}=\frac{g_{n}\sqrt{\kappa}a_{in}}{\kappa+i\Delta_{c}},\label{noise}
\end{equation}
in matrix $A_n$. For TLS's to exhibit quantum coherence, the decoherence rates $\bar{\gamma}_{1,2}$ must be weaker than the other time-scales in the system, as will be discussed below. The effective coupling constant can be derived from $B_{n}$ with
\begin{equation}
  \lambda_{nm}=\frac{-ig_{n}g_{m}}{\kappa+i\Delta_{c}}
  \label{eq:c2}
\end{equation}
and satisfies $\lambda_{nm}=\lambda_{mn}^{\star}$. The coupling depends on $\Delta_{c}$ in a similar way as does the Rabi frequency $\Omega_{n}$. 

The following effective Hamiltonian can then be derived for the TLS's:
\begin{eqnarray}
   \widetilde{H}_{eff} & = & \sum\frac{\bar{\Delta}_{n}}{2}\sigma_{nz}+\Omega_{n}\sigma_{+}+ 
                         \Omega_{n}^{\star}\sigma_{-}\label{eq:Heff}\\
               & + & \sum_{\langle n,m\rangle}\lambda_{nm}\sigma_{n+}\sigma_{m-}+
                         \lambda_{nm}^{\star}\sigma_{m+}\sigma_{n-}\nonumber.
\end{eqnarray}
An interesting difference between the coupling in Eq. (\ref{eq:c1}) and in Eq. (\ref{eq:c2}) is that the effective coupling in Eq. (\ref{eq:c2})  doesn't depend on the frequencies of the TLS's.
\begin{figure}
\begin{center}
\includegraphics[width=6.5cm,clip]{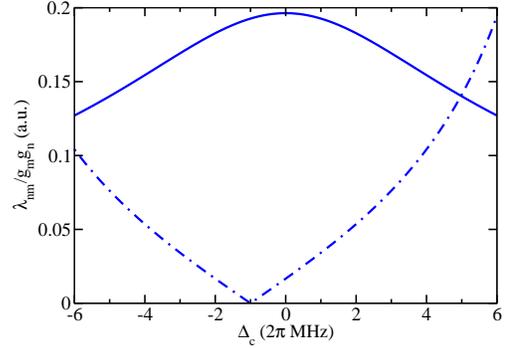}
\end{center}
\caption{Couplings versus detuning. Dash-dotted line: coupling in Eq. (\ref{eq:c1}) and solid line:  coupling in Eq. (\ref{eq:c2}). The parameters are $\Delta_{n}=2\pi\times10\,\mbox{MHz}$, $\Delta_{m} = 2\pi\times32\,\mbox{MHz}$, and $\kappa=2\pi\times5.1\,\mbox{MHz}$.}
\label{fig2}
\end{figure}
In Fig.\ref{fig2}, we plot the magnitude of these two couplings for comparison. 
 
\section{Quantum Logic Operations}
The above system can be used to implement universal quantum operations on the TLS's. The driving amplitude, driving frequency, and the resonator frequency $\omega_{c}$ (or detuning $\Delta_{c}$) can be independently adjusted. Choosing two linearly-independent Hamiltonians $H_{1}$ and $H_{2}$ from the general expression of $\widetilde{H}_{eff}$, a complete set of operators can be constructed from the commutators such as $[H_{1},H_{2}]$, $[H_{1},[H_{1},H_{2}]]$, \emph{etc.}~\cite{LloydPRL1995}. This shows that universal quantum gates can be realized by adjusting the above parameters. Details on how to realized the gates such as the SWAP gate and the Hadamard gate can be found elsewhere~\cite{TianPreprint}.
 
The coupling between the TLS's and the resonator induces extra decoherence on the TLS's due to the decay of the junction resonator. For coupling with a high-Q resonator, a  noise term $\widetilde{H}_{\kappa}$ is generated; for coupling with a strongly-damped resonator, a term $\Lambda_n$ is generated. To realize quantum logic gates, it is necessary that the decoherence rates are much smaller than the effective Rabi frequency $\Omega_{nx}$ and the effective coupling constant $\lambda_{mn}$. It can be shown that the decoherence rate for coupling with a high-Q resonator is $\tau_d^{-1}\sim g_n^2\kappa/\Delta_{nc}^2$. With $\kappa\ll |\Delta_{nc}|$, hundreds of quantum operations can be performed within the decoherence time even when the decay rate of a resonator is a few megahertz, as has been studied in detail in Ref.~\cite{TianPreprint}. However, the decoherence rate for coupling with a strongly-damped resonator is $\tau_d^{-1}\sim g_n^2\kappa/(\kappa^2+\Delta_c^2)$. With $\kappa\ge\Delta_c$, we have $\tau_d^{-1}\sim \Omega_{nx}, \lambda_{mn}$,  and only a few operations can be performed at the best.

\section{Readout}
The junction resonator can function as a readout device for the TLS's. We consider the measurement of the $n^{\mbox{\scriptsize th}}$ TLS  in the dispersive regime.  Let the frequency of the resonator be close to the frequency of this TLS, but have an off-resonance that satisfies $g_n\ll |\Delta_{nc}|$. All the other TLS's are very far off-resonance from the resonator, and so have much smaller Stark shifts. This can be achieved by switching the resonator frequency and the driving frequency at a nanosecond time-scale.  Because the dynamics of the TLS's happens at a much slower rate than this switching rate, the state of the TLS can be considered unaffected during the switching. A measurement of the transmission or reflection in the junction resonator can be used to reveal the qubit states. 

Meanwhile, phase sensitive detection of the stationary state of the resonator can give direct measurement of  the TLS's for a strongly-damped resonator, according to Eq. (\ref{eq:as}). For example, we have 
\begin{equation}
a+a^{\dagger}=\frac{-2\epsilon_{0}\Delta_{c}}{\kappa^{2}+\Delta_{c}^{2}}-\frac{\sum_ng_{n}\kappa\sigma_{ny}+g_{n}\Delta_{c}\sigma_{nx}}{\kappa^{2}+\Delta_{c}^{2}}.
\end{equation}
When the couplings ($g_{n}$'s) are different for different TLS's, the output of the resonator provides a readout of multiple TLS's  in a single measurement~\cite{BlaisPRA2004}.  In addition, by choosing the phase of the measured canonical variable of the resonator, we can choose which TLS operator will be measured. With $\phi=\arg(-ig_{1}/(\kappa+i\Delta_{c}))$, we have $ae^{-i\phi}+a^{\dagger}e^{i\phi} \propto\sigma_{1x}$ and a measurement  of $\sigma_{1x}$ is performed.

This scheme also provides a measurement of the time dependence of the properties of the TLS's. It can be shown that
\begin{equation}
 \langle a^{\dagger}(t)a(t)\rangle=\textrm{Tr}_{s,r}[e^{i\bar{H}_{t}t}a^{\dagger}e^{-i\bar{H}_{t}t}aW(0)]\label{det}\end{equation}
where $W(0)$ is the initial density matrix including the environmental degrees of freedom and $\bar{H}_{t}$ is the total Hamiltonian including the system and the bath. The trace is taken over the system modes (index $s$) and the bath modes (index $r$). Substituting the expression in Eq. (\ref{eq:as}) into Eq. (\ref{det}) and considering one TLS ($n=1$) for simplicity, we derive that
\begin{equation}
\langle a^{\dagger}(t)a(t)\rangle=\frac{g_{1}^{2}C(t)+g_{1}\epsilon_{0}M(t)+\epsilon_{0}^{2}}{\kappa^{2}+\Delta_{c}^{2}}
\end{equation}
where $C(t)=\langle\sigma_{1+}(t)\sigma_{1-}(t)\rangle$ and $M(t)=\langle\sigma_{1+}(t)+\sigma_{1-}(t)\rangle$. The measured results for the correlation function $C(t)$ can be used to interpret the parameters in the matrix $A_{n}$ to reveal the properties of the TLS's via the quantum regression theorem~\cite{WallsBook1994}.

\section{Conclusions}
To conclude, we studied the effective coupling between the TLS's inside a Josephson junction using quantum optics approaches.  Two situations are studied and compared:  TLS's coupling with a high-Q resonator mode and TLS's coupling with a strongly-damped resonator mode.  Our results indicate that the couplings in these two regime have very different properties. Universal quantum gates can be realized on the TLS's when they are  coupled with a high-Q resonator. While the fast decay of a strongly-damped resonator can destroy the coherence of the qubits and affect the successful realization of the quantum gates. We also discussed the readout of the TLS qubits  via the detection of the junction resonator.

\end{document}